\documentclass[brazil,english]{article}
\usepackage[latin9]{inputenc}
\usepackage{geometry}
\usepackage{amsmath}
\usepackage{amssymb}
\usepackage{babel}

\begin{document}

\title{Photon propagation in noncommutative QED with constant external field}
\author{R. Fresneda$^{1}$\thanks{%
Electronic address: rodrigo.fresneda@ufabc.edu.br},\ D.M. Gitman$^{2,3,4}$%
\thanks{%
Electronic address: gitman@dfn.if.usp.br} \ and A.E. Shabad$^{3,2}$\thanks{%
Electronic address: shabad@lpi.ru} \\
$^{1}$\textsl{Federal University of ABC (UFABC), Av. dos Estados, 5001,}\\
\textsl{\ Bairro\ Bangu, CEP 09210-580, Santo André, S. P., Brazil }\\
$^{2}$\textsl{P. Tomsk State University, Lenin Prospekt 36, Tomsk 634050,
Russia} \\
\textsl{Caixa Postal 66318, CEP 05508-090, São Paulo, S. P., Brazil} \\
$^{3}$\textsl{P. N. Lebedev Physics Institute, Leninsky Prospekt 53, Moscow
\ 117924, Russia} \\
$^{4}$\textsl{Instituto de Física, Universidade de São Paulo, Caixa Postal
66318, CEP 05508-090, São Paulo, S. P., Brazil}}
\maketitle

\begin{abstract}
We find dispersion laws for the photon propagating in the presence of
mutually orthogonal constant external electric and magnetic fields in the
context of the $\theta $-expanded noncommutative QED. We show that there is
no birefringence to the first order in the noncommutativity parameter $%
\theta .$ By analyzing the group velocities of the photon eigenmodes we show
that there occurs superluminal propagation for any direction. This
phenomenon depends on the mutual orientation of the external electromagnetic
fields and the noncommutativity vector. We argue that the propagation of
signals with superluminal group velocity violates causality in spite of the
fact that the noncommutative theory is not Lorentz-invariant and speculate
about possible workarounds.
\end{abstract}

\section{Introduction}

The general belief \cite{DFR} that space-time at the Plank scale is
quantized has materialized in the abundant development over the last decades
in the field of noncommutative quantum field theory and quantum mechanics 
\cite{nekrasov}. Despite the general failure of the hope to regularize
quantum field theory \cite{minwalla,belov,chepelev} and the scarcity of
renormalizable models \cite{GW,Fresneda:2008sr}, this field is still being
actively researched \cite{Aga, WanHuaShe}. With regard to the usual
realization of noncommutativity in physics, where the star product in the
deformed algebra of functions is given in terms of a constant antisymmetric
matrix, numerous phenomenological consequences from noncommutativity have
placed stringent conditions on the magnitude of the noncommutativity
parameter \cite{Aga, Horvat, AdoGitShabVas}.

Noncommutative theories constitute an example of theories with Lorentz
symmetry violation, since they contain an \textquotedblleft
external\textquotedblright\ antisymmetric tensor $\theta _{\mu \nu }$
stemming from the commutation relation between the operator-valued
coordinates $[X_{\mu },X_{\nu }]=i$ $\theta _{\mu \nu },$ and should be
considered in this context \cite{Kostelecky}. To theories with broken
relativistic invariance, superluminal propagation is generally peculiar,
starting with the theoretical evidence in Ref. \cite{drumhath} of photons
that propagate faster than light in the background, specially Schwarzschild,
metrics. Another context for superluminal propagation is presented by
quantum electrodynamics, when its Lorentz-invariance is violated by the
presence of an exponentially high external (\textit{e.g.}, magnetic) field 
\cite{ShabUs2011}, owing to the lack of asymptotic freedom in that theory.
There is a vast literature where the superluminal signals in
Lorentz-violated theories \cite{dolgovnovikov} are revealed and discussed 
\footnote{%
Referring to the speed of the wave-front propagation following \cite{brill}.
We consider the group velocity also as that of a signal following Ref. \cite%
{bornwolf, landlif} . It must not exceed the speed of light. As for the
phase velocity, its use by some authors as a criterion for superluminosity
is not justified: it is permitted to outrun light without any contradiction
with causality.}.

The dramatic role of the superluminal propagation in destroying the
causality principle is often underestimated, following the view that once
there is no relativistic invariance from the outset, one may not be upset by
the appearance of a superluminal signal. Such an attitude is, certainly, too
thoughtless. The simplest refutation may be found in the example of the
electrodynamics in a medium, whose presence violates Lorentz-invariance, --
formally by the involvement of an external vector of the 4-velocity of the
medium. The group velocity of an electromagnetic wave in a moving medium is
its group velocity in the medium at rest relativistically added to the speed
of the medium. (This statement also can be extended to a
Lorentz-non-invariant vacuum, characterized by the presence of any external
vector or a tensor \cite{to be published} ). If a superluminal signal should
exist in the medium at rest, it would also be superluminal in the moving
frame and might lead to the paradoxical reversal of the time coordinates of
time-like separated events, following the same standard consideration of a
Lorentz-invariant vacuum.

In the present paper we are dealing with the problem of propagation of
electromagnetic waves when an external constant electromagnetic field is
present, in the nonlinear electrodynamics to which the noncommutative theory
with the Abelian gauge group is reduced by the Seiberg-Witten map \cite%
{Seiberg1999}. In the lowest order of noncommutativity, the nonlinearity is
only cubic in the electromagnetic field. We restrict ourselves to the
so-called space-space noncommutativity, when the noncommutativity tensor has
its time components vanishing in a certain reference frame, and we are
working within the lowest nontrivial term in the expansion of the Taylor
series in powers of the noncommutativity parameter. We impose mutually
orthogonal electric and magnetic fields of arbitrary magnitudes, but
constant in time and space (the property of orthogonality is retained in any
Lorentz frame). Unlike the Lorentz-invariant vacuum, neither of these fields
can be excluded by a Lorentz boost, since this boost would change the
noncommutativity parameter (by supplying time components to it). Therefore,
our case is more general, indeed, than the case of the magnetic field alone,
considered previously in the literature \cite{Guralnik2001}, because it
includes both fields, but it does not overlap with the external field of
Ref. \cite{cai}, where electric and magnetic fields are taken together, but
restricted to the condition that -- in the frame where the noncommutativity
tensor has vanishing time components -- they are mutually parallel and
parallel to the noncommutativity (pseudo)vector \footnote{%
In that paper the author first points the superluminal propagation in the
direction perpendicular to the fields.}.

In the most general constant external field, we calculate the second- and
third-rank polarization tensors (Section 2), responsible, respectively, for
the light propagation and for the electromagnetic wave splitting into two.
We consider the general problem of light propagation by finding the
eigenvalues of the second-rank polarization tensor in Sections 3 and 5. The
photon dispersion laws are found for the simplest special case, where a
magnetic or electric field is parallel to the noncommutativity
(pseudo)vector $\theta _{i}=\epsilon _{ijk}\theta _{jk}$ in Section 4. The
dispersion laws for mutually perpendicular electric and magnetic fields are
established in Section 6. Also the two intermediate special cases, where
there is either only magnetic or only electric external fields are
considered. In all cases the absence of birefringence is noted and formulas
for the group velocities are derived. In all cases the group velocities
proves to exceed the speed of light $c$, taken as unity, when certain
relations between the directions of the external fields and the
noncommutative vector take place. While in the case where only electric or
magnetic fields are present, a direction of propagation (namely, the one
along the coinciding directions of the fields and $\mathbf{\theta })$
exists, where the speed of propagation is simply unity, no such direction is
found in the general context of Section 6, where both fields are present
simultaneously.

In the concluding remarks of Section 7 we discuss two possible scenarios
intended to avoid the verdict -- that suggests itself -- on the
inconsistency of the noncommutative theory due to the incompatibility with
the causality principle.

\section{ Inclusion of an external field in the noncommutative Maxwell action%
}

Following \cite{Bichl:2001nf,Guralnik2001} we consider the first-order
Seiberg-Witten map \cite{Seiberg1999} of the noncommutative Maxwell theory
that results in the action for the electromagnetic field $F_{\mu \nu }$ 
\begin{eqnarray}
S_{SW} &=&-\frac{1}{4}\int d^{4}xF^{\mu \nu }F_{\mu \nu }\text{ }  \notag \\
&&-\frac{1}{2}\int d^{4}x\theta ^{\alpha \beta }F^{\mu \nu }F_{\alpha \mu
}F_{\beta \nu }+\frac{1}{8}\int d^{4}x\theta ^{\alpha \beta }F_{\alpha \beta
}F^{\mu \nu }F_{\mu \nu }+O\left( \theta ^{2}\right) .  \label{eq:sw-action}
\end{eqnarray}%
It is understood that the coupling constant is included in $\theta$.
Throughout this paper we restrict ourselves to space-space noncommutativity.
That is, in Lorentz invariant terms, we require the following relations 
\footnote{%
Here the contraction of the electromagnetic field strength tensor $F_{\mu
\nu }$ with its dual $\tilde{F}_{\mu \nu }$ is to be understood as $F\tilde{F%
}=F_{\mu \nu }\tilde{F}^{\nu \mu }$. Likewise, $\theta ^{2}=\theta _{\mu \nu
}\theta ^{\nu \mu }$ and $\left( \theta \tilde{\theta}\right) _{\mu \nu
}\equiv \theta _{\mu \alpha }\tilde{\theta}_{\,\,\nu }^{\alpha }$. For a
summary of our notational conventions, see the Appendix.} 
\begin{equation}
(\theta \tilde{\theta})_{\mu \nu }=0,\qquad \theta ^{2}<0,  \label{theta^2}
\end{equation}
to be obeyed by the noncommutativity tensor, which implyZ the existence of a
Lorentz frame - the special frame - where the noncommutativity tensor only
has space-space components: $\theta _{0i}=0$.

In what follows, it is convenient to divide the Lagrangian density from (\ref%
{eq:sw-action}) in two parts, 
\begin{equation}
\mathcal{L}_{SW}=\mathcal{L}_{0}+\mathfrak{L}\,,  \label{eq:lagrange-density}
\end{equation}%
where 
\begin{equation*}
\mathcal{L}_{0}=-\frac{1}{4}F^{\mu \nu }F_{\mu \nu }\,,\,\,\mathfrak{L}=-%
\frac{1}{2}\theta ^{\alpha \beta }F^{\mu \nu }F_{\alpha \mu }F_{\beta \nu }+%
\frac{1}{8}\theta ^{\alpha \beta }F_{\alpha \beta }F^{\mu \nu }F_{\mu \nu }.
\end{equation*}

Let us divide the gauge connection into two parts, a dynamic field $a$ and
an external field $\mathcal{A}$, 
\begin{equation*}
A_{\mu }=a_{\mu }+\mathcal{A}_{\mu }\,.
\end{equation*}%
Then, 
\begin{equation*}
F_{\mu \nu }=f_{\mu \nu }+\mathcal{F}_{\mu \nu }\,,\,\,f_{\mu \nu }=\partial
_{\mu }a_{\nu }-\partial _{\nu }a_{\mu }\,,\,\,\mathcal{F}_{\mu \nu
}=\partial _{\mu }\mathcal{A}_{\nu }-\partial _{\nu }\mathcal{A}_{\mu }
\end{equation*}%
and the action $S_{SW}$ (\ref{eq:sw-action}) becomes 
\begin{eqnarray*}
&&S_{SW}=-\frac{1}{4}\int d^{4}x\left( f^{\mu \nu }f_{\mu \nu }+\mathcal{F}%
^{\mu \nu }\mathcal{F}_{\mu \nu }+2f^{\mu \nu }\mathcal{F}_{\mu \nu }\right)
\\
&&-\frac{1}{2}\int d^{4}x\theta ^{\alpha \beta }\left( f^{\mu \nu }f_{\alpha
\mu }f_{\beta \nu }+\mathcal{F}^{\mu \nu }\mathcal{F}_{\alpha \mu }\mathcal{F%
}_{\beta \nu }+2f^{\mu \nu }f_{\alpha \mu }\mathcal{F}_{\beta \nu }\right. \\
&&\left. +f^{\mu \nu }\mathcal{F}_{\alpha \mu }\mathcal{F}_{\beta \nu }+%
\mathcal{F}^{\mu \nu }f_{\alpha \mu }f_{\beta \nu }+2\mathcal{F}^{\mu \nu
}f_{\alpha \mu }\mathcal{F}_{\beta \nu }\right) \\
&&+\frac{1}{8}\int d^{4}x\theta ^{\alpha \beta }f_{\alpha \beta }\left(
f^{\mu \nu }f_{\mu \nu }+\mathcal{F}^{\mu \nu }\mathcal{F}_{\mu \nu
}+2f^{\mu \nu }\mathcal{F}_{\mu \nu }\right) \\
&&+\frac{1}{8}\int d^{4}x\theta ^{\alpha \beta }\mathcal{F}_{\alpha \beta
}\left( f^{\mu \nu }f_{\mu \nu }+\mathcal{F}^{\mu \nu }\mathcal{F}_{\mu \nu
}+2f^{\mu \nu }\mathcal{F}_{\mu \nu }\right) \,.
\end{eqnarray*}%
Let us write the integrand here as 
\begin{equation}
\mathcal{L}_{SW}=\mathcal{L}\left( \mathcal{F}\right) +\frac{1}{2}D_{\mu \nu
}^{-1}a^{\mu }a^{\nu }+\frac{1}{6}\Pi _{\mu \nu \sigma }a^{\mu }a^{\nu
}a^{\sigma }\,,  \label{eq:a-expansion}
\end{equation}%
where 
\begin{equation}
D_{\mu \nu }^{-1}=k^{2}\eta _{\mu \nu }-k_{\mu }k_{\nu }+\Pi _{\mu \nu }\,
\label{eq:inverse-propagator}
\end{equation}%
is the photon propagator. The polarization second- and third-rank tensors
are defined as 
\begin{equation}
\Pi _{\mu \nu }=\left. \frac{\partial ^{2}\mathfrak{L}}{\partial A_{\mu
}\partial A_{\nu }}\right\vert _{F=\mathcal{F}},\qquad \Pi _{\mu \nu \rho
}=\left. \frac{\partial ^{3}L}{\partial A_{\mu }\partial A_{\nu }\partial
A_{\rho }}\right\vert _{F=\mathcal{F}}.  \label{tensors}
\end{equation}%
These are transverse in every index: $\Pi _{\mu \nu }k_{\nu }=\Pi _{\mu \nu
\rho }k_{\rho }=0$, since $\mathcal{L}$ depends only on the field strength.

The first term on the right-hand side of (\ref{eq:a-expansion}), $\mathcal{L}%
\left( \mathcal{F}\right) $, is 
\begin{equation}
\mathcal{L}\left( \mathcal{F}\right) =-\frac{1}{4}\mathcal{F}^{\mu \nu }%
\mathcal{F}_{\mu \nu }-\frac{1}{2}\theta ^{\alpha \beta }\mathcal{F}^{\mu
\nu }\mathcal{F}_{\alpha \mu }\mathcal{F}_{\beta \nu }+\frac{1}{8}\theta
^{\alpha \beta }\mathcal{F}_{\alpha \beta }\mathcal{F}^{\mu \nu }\mathcal{F}%
_{\mu \nu }\,.  \label{eq:external-lagrangian}
\end{equation}

We did not include the term linear in $a$ in (\ref{eq:a-expansion}), since
it vanishes for constant external fields $\mathcal{F}_{\mu \nu }=const$ ,
because $\left. \frac{\partial \mathcal{L}_{SW}}{\partial A_{\mu }}%
\right\vert _{A=\mathcal{A}}=$ $\frac{\partial }{\partial x_{\nu }}\left. 
\frac{\partial \mathcal{L}_{SW}}{\partial F_{\nu \mu }}\right\vert _{F=%
\mathcal{F}}=0.$ In other words, any constant field, in no way correlated
with the tensor $\theta _{\alpha \beta }$, is an exact source-free solution
to the equation of motion.

The part quadratic in the dynamic fields $a^{\mu }$ contains the terms 
\begin{equation*}
-\frac{1}{4}f^{\mu \nu }f_{\mu \nu }-\theta ^{\alpha \beta }f^{\mu \nu
}f_{\alpha \mu }\mathcal{F}_{\beta \nu }-\frac{1}{2}\theta ^{\alpha \beta }%
\mathcal{F}^{\mu \nu }f_{\alpha \mu }f_{\beta \nu }+\frac{1}{4}\theta
^{\alpha \beta }f_{\alpha \beta }f^{\mu \nu }\mathcal{F}_{\mu \nu }+\frac{1}{%
8}\theta ^{\alpha \beta }\mathcal{F}_{\alpha \beta }f^{\mu \nu }f_{\mu \nu
}\,,
\end{equation*}%
while cubic contributions come from 
\begin{equation*}
-\frac{1}{2}\theta ^{\alpha \beta }f^{\mu \nu }f_{\alpha \mu }f_{\beta \nu }+%
\frac{1}{8}\theta ^{\alpha \beta }f_{\alpha \beta }f^{\mu \nu }f_{\mu \nu
}\,.
\end{equation*}%
Taking into account that the Fourier transform of $\partial _{\mu }a_{\nu
}\left( x\right) $ is $ik_{\mu }a_{\nu }\left( k\right) $, the quadratic
contribution gives rise to the expression for the photon propagator 
\begin{align}
D_{\mu \nu }^{-1}& =k^{2}\eta _{\mu \nu }-k_{\mu }k_{\nu }+\frac{1}{2}\left(
\theta \mathcal{F}\right) \left( k^{2}\eta _{\mu \nu }-k_{\mu }k_{\nu
}\right)  \notag \\
& -\left\{ \left[ \left( \theta \mathcal{F}\right) _{\mu \nu }+\left( \theta 
\mathcal{F}\right) _{\nu \mu }\right] k^{2}+2\left( k\theta \mathcal{F}%
k\right) \eta _{\mu \nu }-\left( k\left[ \theta \mathcal{F}+\mathcal{F}%
\theta \right] \right) _{\nu }k_{\mu }-\left( k\left[ \mathcal{F}\theta
+\theta \mathcal{F}\right] \right) _{\mu }k_{\nu }\right\}  \label{D}
\end{align}%
and to the polarization tensor 
\begin{align}
\Pi _{\mu \nu }& =\frac{1}{2}\left( \theta \mathcal{F}\right) \left(
k^{2}\eta _{\mu \nu }-k_{\mu }k_{\nu }\right)  \notag \\
& -\left[ \left( \theta \mathcal{F}\right) _{\mu \nu }+\left( \theta 
\mathcal{F}\right) _{\nu \mu }\right] k^{2}-2\left( k\theta \mathcal{F}%
k\right) \eta _{\mu \nu }+\left( k\left[ \theta \mathcal{F}+\mathcal{F}%
\theta \right] \right) _{\nu }k_{\mu }+\left( k\left[ \mathcal{F}\theta
+\theta \mathcal{F}\right] \right) _{\mu }k_{\nu }\,,
\label{eq:polarization-tensor}
\end{align}%
quadratic in $k$. The third-rank polarization tensor is given by the
expression

\begin{equation*}
\Pi _{\mu \nu \sigma }=3ik^{\alpha }\theta _{\alpha \sigma }\left( k^{2}\eta
_{\mu \nu }-k_{\mu }k_{\nu }\right) ,
\end{equation*}%
cubic in $k$. This character of dependence on the 4-momentum vector is a
direct consequence of the fact that the action (\ref{eq:sw-action}) is local
in the sense that it does not include space and time derivatives of the
field intensities, the same as in the local or infrared, $k\rightarrow 0,$
approximation of quantum electrodynamics with or without a constant external
field, in which case the power of $k$ coincides with the rank of the
polarization tensor \cite{AdoCosGitSha}. The equation of motion for the
field $a$ above the external field background, resulting from (\ref%
{eq:a-expansion}), is the classical quadratic equation of nonlinear optics 
\begin{equation*}
\frac{\partial \mathcal{L}}{\partial A_{\mu }}=\Pi _{\mu \nu }a_{\mu }+\frac{%
1}{2}\Pi _{\mu \nu \rho }a_{\nu }a_{\rho }=0.
\end{equation*}%
It describes in a classical way a transformation of a photon into two (or 
\textit{vice versa}), which may be also treated via classical scattering
theory in probabilistic terms after we define one- or two-photon asymptotic
states. The probabilities of photon splitting/merging are given in terms of
the third-rank polarization tensor, whereas the photon normal modes, as well
as integrals of motion associated with conserved Noether generators of
canonical transformations within the Hamiltonian formalism are determined by
the second-rank polarization tensor. Therefore the term quadratic in $a$ of (%
\ref{eq:a-expansion}) is taken as the \textquotedblleft
free\textquotedblright\ Lagrangian. Then the cubic term should be treated as
the \textquotedblleft interaction\textquotedblright\ part.

\section{General covariant description of photon propagation}

Since the external field and the noncommutativity parameter $\theta $ are
constant, the second rank polarization tensor only depends on coordinate
differences and hence on one momentum $k_{\mu }$. By calculating the second
derivative of $\mathcal{L}_{SW}$ with respect to the potential, we arrive at
the expression (\ref{eq:polarization-tensor}), which is a linear combination
of symmetric transverse matrices. The symmetry is explicitly provided by the
relation $(\mathcal{F}\theta )_{\mu \nu }=(\theta \mathcal{F})_{\nu \mu }$
that follows from the antisymmetry of the matrices $\theta $ and $\mathcal{F}
$.

It is noteworthy that when the external field is absent, the polarization
operator is zero in spite of the presence of the noncommutativity parameter $%
\theta $. Thus, the dispersion laws are in that case all trivial $%
k^{2}=k^{2}-k_{0}^{2}=0$.

The eigenvalues of the vacuum polarization tensor define the energy spectrum
of free electromagnetic waves propagating in the external field. The most
general gauge-invariant expression for the polarization tensor in a constant
and homogeneous background (including an electromagnetic field and the
noncommutativity tensor) is \cite{shabtrudy} 
\begin{equation}
\Pi _{\mu \nu }\left( k\right) =\sum_{i=1}^{6}\Pi _{i}\left( I\right) \Psi
_{\mu \nu }^{\left( i\right) }\,,  \label{eq:gen-gauge-inv-exp}
\end{equation}%
where $\Psi ^{\left( i\right) }$ are linearly independent transverse
symmetric matrices, and the coefficients $\Pi _{i}\left( I\right) $ depend
on the invariants of the theory, such as the ones involving the vector $%
k_{\mu }$ and the electromagnetic field strength $\mathcal{F}_{\mu \nu }$, $%
k^{2},k\mathcal{F}^{2}k$, $\mathfrak{F}$ and $\mathfrak{G}$, and also the
ones depending on the external tensor $\theta _{\mu \nu }$, such as $k%
\mathcal{F}^{2}\theta \mathcal{F}k$. The number of independent matrices in (%
\ref{eq:gen-gauge-inv-exp}) is 6, because the symmetry conditions $\Pi
_{\mu\nu }\left( k\right) =\Pi _{\nu \mu }\left( k\right) $ following from
the definition \ref{tensors} leave $10$ out of $16= 4 \times 4$ independent
components of the polarization operator, whereas the four transversality
conditions $\Pi _{\mu \nu }\left( k\right) k^{\nu }=0$ reduce their number
to $6$.

In general, in order to find the Green function $D$ as well as dispersion
laws for photon eigenmodes it is necessary to diagonalize the polarization
tensor (\ref{eq:polarization-tensor}). The polarization operator has three
nontrivial scalar eigenvalues $\varkappa _{i}$, $i=1,2,3$ ($\varkappa _{4}=0$
due to its transversality) 
\begin{equation}
\Pi _{\mu }^{\,\,\,\nu }d_{\nu }^{\left( i\right) }=\varkappa _{i}d_{\mu
}^{\left( i\right) }\,,\,i=1,2,3,4\,  \label{eigen}
\end{equation}%
(that may depend on the scalars available in the problem), which are linear
in $\theta $ and $\mathcal{F}$, and quadratic in $k$. Then the dispersion
equations are 
\begin{equation}
\varkappa _{i}=k^{2},\text{ \ \ \ \ }i=1,2,3\,,  \label{disp}
\end{equation}%
while the three eigenvectors $d_{\mu }^{\left( i\right) }$ carry information
about polarizations of the three (as a matter of fact -- two) eigenmodes. To
find the eigenvalues and eigenvectors one generally needs to solve a
cumbersome cubic equation. More important is that the corresponding
eigenvectors are not universal, but depend on dynamics, that is on $\theta$,
which includes the coupling constant.

We shall proceed with the general case in Section 5, where this cubic
equation will be avoided by considering only the lowest order in $\theta$.
But preliminarily, in the next subsection, we consider the special case when
the magnetic field is parallel to the pseudovector $\theta _{i}=\varepsilon
_{ijk}\theta _{jk}$, in a frame where $\theta _{0i}=0$. In this case, the
diagonalization of the polarization tensor simplifies to a closed-form
purely kinematic solution, the same as in the standard, commutative
electrodynamics in external magnetic field, \cite{shabtrudy}.

\section{ Magnetic field parallel to the noncommutativity pseudovector $%
\boldsymbol{\protect\theta }$}

In the special case to be considered now the simplest relation between the
external field and the noncommutativity tensor $\theta _{\mu \nu }=\frac{1}{2%
}\alpha \mathcal{F}_{\mu \nu }$ is adopted, where $\alpha $ is a dimensional
parameter ($\left[ \alpha \right] =mass^{-4}$) . Consequently, in view of (%
\ref{theta^2}), also $\mathfrak{G}=0$, $(k\theta \widetilde{\mathcal{F}}k)=0$%
, $\mathfrak{F}>0$. As a result, the external field is purely magnetic in
the special frame, $\mathcal{B}_{i}=\frac{1}{2}\epsilon _{ijk}\mathcal{F}%
_{jk}$, and parallel to the noncommutativity pseudovector $\boldsymbol{%
\theta }$, defined as $\theta _{i}=\epsilon _{ijk}\theta _{jk}$. We
therefore align the axes of our special frame so that the third axis
coincides with the direction of the magnetic field. Then, $\theta
_{1}=\theta _{2}=\mathcal{B}_{1}=\mathcal{B}_{2}=0$, in other words, $\theta
_{3}=$ $\theta _{12}$ and $\mathcal{F}_{12}$ are the only nonvanishing
components of the noncommutativity vector and of the external field strength
tensor.

With the condition $\theta _{\mu \nu }=\frac{1}{2}\alpha \mathcal{F}_{\mu
\nu }$ the field tensor remains the only external tensor in the problem, and
the covariant expansion of the polarization operator is the same as in the
problem of commutative electrodynamics in a constant field of the most
general form. While $\mathfrak{G}\neq 0$, the expansion (\ref%
{eq:gen-gauge-inv-exp}) contains four independent matrices $\Psi _{\mu \nu
}^{\left( i\right) }$, given by \cite{shabtrudy} 
\begin{align}
\Psi _{\mu \nu }^{(1)}=& k^{2}\eta _{\mu \nu }-k_{\mu }k_{\nu }\,,\,\,\Psi
_{\mu \nu }^{(2)}=-(\mathcal{F}k)_{\mu }(\mathcal{F}k)_{\nu }\,,\Psi _{\mu
\nu }^{(3)}=-k^{2}\left( \delta _{\mu }^{\sigma }-\frac{k_{\mu }k^{\sigma }}{%
k^{2}}\right) F_{\sigma \kappa }^{2}\left( \delta _{\nu }^{\kappa }-\frac{%
k^{\kappa }k_{\nu }}{k^{2}}\right) ,  \notag \\
& \Psi _{\mu \nu }^{\left( 4\right) }=\left( \mathcal{F}k\right) _{\mu }%
\mathcal{F}_{\nu \sigma }^{3}k^{\sigma }+\mathcal{F}_{\mu \sigma
}^{3}k^{\sigma }\left( \mathcal{F}k\right) _{\nu }.  \label{eq:psi-matrices}
\end{align}%
In our present special case $\mathfrak{G}=0$, therefore one has $\mathcal{F}%
_{\mu \nu }^{3}=-2\mathfrak{F}\mathcal{F_{\mu \nu }}$. As a result, the set
of matrices $\Psi ^{\left( i\right) }$ is no longer linear independent ($%
\Psi ^{\left( 4\right) }$ is proportional to $\Psi ^{\left( 2\right) }$).
Moreover, the eigenvectors $d_{\mu }^{\left( i\right) }$, $i=1,2,3,4$, of
the polarization tensor have the simple form $d_{\mu }^{\left( i\right)
}=b_{\mu }^{\left( i\right) }$ 
\begin{equation}
b_{\mu }^{\left( 1\right) }=\left( \mathcal{F}^{2}k\right) _{\mu
}k^{2}-k_{\mu }\left( k\mathcal{F}^{2}k\right) \,,\,\,b_{\mu }^{\left(
2\right) }=(\tilde{\mathcal{F}}k)_{\mu }\,,\,\,b_{\mu }^{\left( 3\right) }=(%
\mathcal{F}k)_{\mu }\,,\,\,b_{\mu }^{(4)}=k_{\mu }\,.
\label{eq:orthogonal-basis}
\end{equation}%
The nonvanishing eigenvalues are given in terms of the coefficients $\Pi
_{i} $ in (\ref{eq:gen-gauge-inv-exp}) as: 
\begin{align}
\varkappa _{1}& =\Pi _{1}k^{2}+\Pi _{3}\left( k\mathcal{F}^{2}k+2\mathfrak{F}%
k^{2}\right)  \notag \\
\varkappa _{2}& =\Pi _{1}k^{2}  \notag \\
\varkappa _{3}& =\Pi _{1}k^{2}+\Pi _{2}k\mathcal{F}^{2}k+\Pi _{3}2\mathfrak{F%
}k^{2}  \label{eq:eigenvalues}
\end{align}

The solutions of the dispersion equations (\ref{disp}) supply poles to the
photon propagator. The equation with $i=1$ has only $k^{2}=0$ as its
solution, which is pure gauge due to the properties of the corresponding
eigenvector $b_{\mu }^{(1)}$.

Using the condition $\theta _{\mu \nu }=\frac{1}{2}\alpha \mathcal{F}_{\mu
\nu }$ we get for the polarization tensor (\ref{eq:polarization-tensor}). 
\begin{align*}
\Pi _{\mu \nu }\left( k\right) & =-\alpha \mathfrak{F}\left( k^{2}\eta _{\mu
\nu }-k_{\mu }k_{\nu }\right) -\alpha \mathcal{F}_{\mu \nu }^{2}k^{2}-\alpha 
\mathcal{F}_{\alpha \beta }^{2}k^{\alpha }k^{\beta }\eta _{\mu \nu }+\alpha
k^{\sigma }\mathcal{F}_{\sigma \nu }^{2}k_{\mu }+\alpha k^{\sigma }\mathcal{F%
}_{\sigma \mu }^{2}k_{\nu } \\
& =-\alpha \left( \mathfrak{F}+\mathcal{F}_{\sigma \kappa }^{2}\frac{%
k^{\sigma }k^{\kappa }}{k^{2}}\right) \left( k^{2}\eta _{\mu \nu }-k_{\mu
}k_{\nu }\right) -\alpha k^{2}\left( \delta _{\mu }^{\sigma }-\frac{%
k^{\sigma }k_{\mu }}{k^{2}}\right) \mathcal{F}_{\sigma \kappa }^{2}\left(
\delta _{\nu }^{\kappa }-\frac{k^{\kappa }k_{\nu }}{k^{2}}\right)
\end{align*}%
Comparing this expression with the general expansion (\ref%
{eq:gen-gauge-inv-exp}), where the $\Psi ^{\left( i\right) }$ matrices are
given by (\ref{eq:psi-matrices}), one finds the coefficients $\Pi _{i}$ to be

\begin{equation*}
\Pi _{1}=-\alpha \mathfrak{F}\left( \frac{k^{2}+\frac{k\mathcal{F}^{2}k}{%
\mathfrak{F}}}{k^{2}}\right) \,,\,\,\Pi _{2}=0\,,\,\,\Pi _{3}=\alpha
\end{equation*}%
Therefore, the nonzero eigenvalues of the polarization tensor are readily
obtained from (\ref{eq:eigenvalues}) taking into account the above
coefficients, 
\begin{equation*}
\varkappa _{1}=\alpha \mathfrak{F}k^{2}\,,\,\,\varkappa _{2}=-\alpha 
\mathfrak{F}\left( k^{2}+\frac{k\mathcal{F}^{2}k}{\mathfrak{F}}\right)
\,,\,\,\varkappa _{3}=\alpha \mathfrak{F}\left( k^{2}-\frac{k\mathcal{F}^{2}k%
}{\mathfrak{F}}\right)
\end{equation*}

The solutions to the equations $\varkappa_{i}=k^{2}$ can be found with the
help of the relations 
\begin{equation*}
k^{2}+\frac{k\mathcal{F}^{2}k}{2\mathfrak{F}}=k_{\parallel}^{2}-k_{0}^{2}\,,%
\,\,\frac{k\mathcal{F}^{2}k}{2\mathfrak{F}}=-k_{\perp}^{2}\,,
\end{equation*}
valid in the special frame. The two-dimensional vector $\mathbf{k}_{\perp}$
is the photon momentum projection onto the plane orthogonal to $\boldsymbol{%
\mathcal{B}}$, while $k_{\parallel}$ is the photon momentum projection onto
the direction of $\boldsymbol{\mathcal{B}}$.

The dispersion equations (\ref{disp}) with $i=2,3$ have the common solution: 
\begin{equation*}
k_{0}^{2}-k_{\parallel }^{2}=\left( 1-\alpha \mathcal{B}_{3}^{2}\right)
k_{\perp }^{2}=\left( 1-2\theta _{3}\mathcal{B}_{3}\right) k_{\perp }^{2}
\end{equation*}%
or 
\begin{equation*}
k_{0}=|\mathbf{k}|-\theta _{3}\mathcal{B}_{3}\frac{k_{\perp }^{2}}{\mathbf{k}%
^{2}}.
\end{equation*}

There is no birefringence (within the linear-in-$\theta $ accuracy adopted),
since these solutions are the same . In this respect the situation is
analogous to the Born-Infeld model of nonlinear electrodynamics, only there
the absence of birefringence is an exact property of the model,
distinguishing it from any other nonlinear electrodynamics.

From the above one can calculate the group velocity $\mathbf{v}_{\mathrm{gr}%
}=\frac{dk_{0}}{d\mathbf{k}},$ whose components across and along the
direction of $\boldsymbol{\mathcal{B}}$ and its norm are, respectively, 
\begin{equation*}
\mathbf{v}_{\mathrm{gr}}^{\perp }=\frac{\mathbf{k}_{\perp }}{|\mathbf{k}|}%
\left( 1-2\theta _{3}\mathcal{B}_{3}+\theta _{3}\mathcal{B}_{3}\frac{\mathbf{%
k}_{\perp }^{2}}{\mathbf{k}^{2}}\right) ,
\end{equation*}

\begin{equation*}
v_{\mathrm{gr}}^{\parallel }=\frac{k_{\parallel }}{|\mathbf{k}|}\left(
1+\theta _{3}\mathcal{B}_{3}\frac{\mathbf{k}_{\perp }^{2}}{\mathbf{k}^{2}}%
\right)
\end{equation*}
\begin{equation}
v_{\mathrm{gr}}=\left\vert \frac{dk_{0}}{d\mathbf{k}}\right\vert =1-\theta
_{3}\mathcal{B}_{3}\frac{\mathbf{k}_{\perp }^{2}}{\mathbf{k}^{2}}.
\label{eq:group-velocity-1}
\end{equation}%
The direction of propagation as identified with that of the group velocity
does not, obviously, coincide with the direction of the wave vector $\mathbf{%
k}$, but makes the angle $\varphi $ with it, such that $\cos \varphi
=1-2\left( \theta _{3}\mathcal{B}_{3}\cos \eta \sin \eta \right) ^{2}$, $%
|\varphi |=|$ $\theta _{3}\mathcal{B}_{3}\sin 2\eta |$, where $\cos \eta =%
\frac{k_{\parallel }}{|\mathbf{k}|}$. The directions of $\mathbf{v}_{\mathrm{%
gr}}$ and $\mathbf{k}$ coincide for propagations parallel $\left( \mathbf{k}%
_{\perp }=0,\text{ }\cos \eta =1\right) $ and perpendicular $\left(
k_{\parallel }=0,\text{ }\cos \eta =0\right) \ $ to the magnetic field.

It is seen from (\ref{eq:group-velocity-1}) that for any direction of
propagation but parallel, the speed can be smaller or larger than one,
depending on whether $\boldsymbol{\mathcal{B}}$ and $\boldsymbol{\theta }$
are parallel or anti-parallel.

\section{General covariant description of photon propagation continued}

In this Section we come back to the general case of Section 3 and proceed by
diagonalization of the inverse propagator (\ref{eq:inverse-propagator}),
instead of diagonalizing the polarization tensor, as in the previous section.

We need to solve the eigenvalue equation for the inverse propagator (\ref{D}%
), which is equivalent to (\ref{eigen}) 
\begin{equation}
\left( D^{-1}\right) _{\mu }^{\,\,\,\sigma }d_{\sigma }=\lambda d_{\mu }
\label{eq:eigenvalue}
\end{equation}%
where the vector $d_{\sigma }$ is a linear combination of the vectors $%
b_{\sigma }^{\left( i\right) }$ (\ref{eq:orthogonal-basis}), which, unlike
the special case of Section 4, no longer are eigenvectors, but still form an
orthogonal basis, 
\begin{equation}
d_{\sigma }=\sum_{i=1}^{4}\alpha _{i}b_{\sigma }^{\left( i\right) }\,.
\label{d}
\end{equation}%
Contrary to the eigenvectors (\ref{eq:orthogonal-basis}) of the problem in
the previous Section, the eigenvectors (\ref{d}) will depend on the
dynamics, i.e. on the noncommutativity parameter that contains the couplinf
constant.

One can check that the basis vectors $b^{\left( i\right) }$ are eigenvectors
of the free part of the propagator, 
\begin{align*}
& \left( \delta _{\mu }^{\sigma }k^{2}-k_{\mu }k^{\sigma }\right) b_{\sigma
}^{\left( a\right) }=k^{2}b_{\mu }^{a}\,,\,\,\mathrm{for}\,\,a=1,2,3 \\
& \left( \delta _{\mu }^{\sigma }k^{2}-k_{\mu }k^{\sigma }\right) b_{\sigma
}^{\left( 4\right) }=0\,.
\end{align*}%
Since the polarization vector is transverse, $\Pi _{\mu }^{\,\,\sigma
}k_{\sigma }=0$, as are the first three basis vectors, equation (\ref%
{eq:eigenvalue}) reduces to 
\begin{equation*}
\left( D^{-1}\right) _{\mu }^{\,\,\,\sigma }\sum_{a=1}^{3}\alpha _{a}b_{\mu
}^{\left( a\right) }=\left( k^{2}-\varkappa \right) \sum_{i=a}^{3}\alpha
_{a}b_{\mu }^{\left( a\right) }\equiv \lambda \sum_{a=1}^{3}\alpha
_{a}b_{\mu }^{\left( a\right) }\,,
\end{equation*}%
where $\varkappa $ is the eigenvalue of the polarization tensor whose
eigenvector is $\sum_{a=1}^{3}\alpha _{a}b_{\mu }^{\left( a\right) }$.
Making use of the orthogonality of the eigenvectors $b^{\left( a\right) }$, $%
b^{\left( a\right) \mu }b_{\mu }^{\left( b\right) }=0$ for $a\neq b$, one
has 
\begin{equation*}
x_{ab}\alpha _{b}\equiv \frac{b^{\left( a\right) \mu }\left( D^{-1}\right)
_{\mu }^{\,\,\,\sigma }b_{\sigma }^{\left( b\right) }}{b^{\left( a\right)
\nu }b_{\nu }^{\left( a\right) }}\alpha _{b}=\lambda \alpha _{a}.
\end{equation*}%
Now in order to determine $\lambda $ we need to solve the equation 
\begin{equation*}
\det \left\vert x_{ab}-\lambda \delta _{ab}\right\vert =0\,,
\end{equation*}%
which expresses $\lambda $ in terms of the matrix elements $x_{ab}$. Taking
into account that the off-diagonal components of $x_{ab}$ are of order $%
\theta $, they only contribute terms of order at least $\theta ^{2}$ in the
characteristic polynomial. Therefore, the dominant contributions come from
the diagonal part, 
\begin{equation*}
\det \left\vert x_{ab}-\lambda \delta _{ab}\right\vert
=\prod_{a=1}^{3}\left( x_{aa}-\lambda \right) +O\left( \theta ^{2}\right) \,.
\end{equation*}%
Therefore, to first order in $\theta $, the dispersion equations (\ref{disp}%
) can be written in the form $\lambda _{a}=x_{aa}=0$.

\section{Mutually orthogonal electric and magnetic fields}

We are going to solve the dispersion relations in the special case where the
electric and magnetic fields are orthogonal, and a Lorentz frame exists
where the noncommutativity tensor has vanishing time-space components .
Therefore, we add the restriction $\mathfrak{G}=0$ to conditions (\ref%
{theta^2}).

In calculating the diagonal components of the matrix $x_{ab}$, the only
nontrivial contribution from the polarization operator (\ref%
{eq:polarization-tensor}) comes from the term $\left( \theta \mathcal{F}%
\right) _{\mu \nu }$, due to the transversality of the basis vectors and the
simplifying relations $\left( \mathcal{F}\tilde{\mathcal{F}}\right) _{\mu
\nu }=0$ and $\mathcal{F}_{\mu \nu }^{3}=-2\mathfrak{F}\mathcal{F_{\mu \nu }}
$, peculiar to the configuration $\mathfrak{G}=0$ (see Appendix). 
\begin{align}
x_{11}& =k^{2}\left( 1-\frac{1}{2}\left( \theta \mathcal{F}\right) \right) +2%
\frac{k^{2}}{\left( k\mathcal{F}^{2}k\right) }\left( k\mathcal{F}^{2}\theta 
\mathcal{F}k\right)  \notag \\
x_{22}& =k^{2}\left( 1-\frac{1}{2}\left( \theta \mathcal{F}\right) \right)
+2\left( k\theta \mathcal{F}k\right)  \notag \\
x_{33}& =k^{2}\left( 1-\frac{1}{2}\left( \theta \mathcal{F}\right) \right)
+2\left( k\theta \mathcal{F}k\right) +2\frac{k^{2}}{\left( k\mathcal{F}%
^{2}k\right) }\left( k\mathcal{F}^{2}\theta \mathcal{F}k\right)
\label{eq:G0eigenvalues}
\end{align}

As in Section 4, we go over to a special reference frame where the
orientation of the spatial axes are such that the third axis is aligned to
the magnetic field, i.e., $\mathcal{B}_{1,2}=0,$ $\mathcal{B}_{3}\neq 0$.
Then the electric field should lie in the (1,2)-plane. Now we may rotate the
spatial frame around the magnetic field (axis 3) to nullify $\theta _{1}$.
Hence the choices $\theta _{2},\theta _{3},\mathcal{B}_{3}\neq 0,\,\mathcal{E%
}_{1,2}\neq 0$ to be considered in the present Section represent the most
general case specified by the conditions (\ref{theta^2}) and $\mathfrak{G}=0$%
.

With the help of the identities (\ref{useful-ids}) from the Appendix, we are
able to write the eigenvalues (\ref{eq:G0eigenvalues}) in the special
reference frame as 
\begin{align*}
x_{11}& =k^{2}\left( 1-\boldsymbol{\theta }\cdot \boldsymbol{\mathcal{B}}%
\right) \\
x_{22}& =k^{2}\left( 1+\boldsymbol{\theta }\cdot \boldsymbol{\mathcal{B}}%
\right) +2\left[ k^{0}\mathbf{k}\cdot \left( \boldsymbol{\theta }\times 
\boldsymbol{\mathcal{E}}\right) -\mathbf{k}^{2}\left( \boldsymbol{\theta }%
\cdot \boldsymbol{\mathcal{B}}\right) +\left( \mathbf{k}\cdot \boldsymbol{%
\mathcal{B}}\right) \left( \mathbf{k}\cdot \boldsymbol{\theta }\right) %
\right] \\
x_{33}& =k^{2}\left( 1-\boldsymbol{\theta }\cdot \boldsymbol{\mathcal{B}}%
\right) +2\left[ k^{0}\mathbf{k}\cdot \left( \boldsymbol{\theta }\times 
\boldsymbol{\mathcal{E}}\right) -\mathbf{k}^{2}\left( \boldsymbol{\theta }%
\cdot \boldsymbol{\mathcal{B}}\right) +\left( \mathbf{k}\cdot \boldsymbol{%
\mathcal{B}}\right) \left( \mathbf{k}\cdot \boldsymbol{\theta }\right) %
\right]
\end{align*}%
As a result, the equation $x_{11}=0$ implies $k^{2}=0$, while the two
equations $x_{22}=x_{33}=0$ imply the common solution 
\begin{equation*}
k^{2}=-2\left[ k^{0}\mathbf{k}\cdot \left( \boldsymbol{\theta }\times 
\boldsymbol{\mathcal{E}}\right) -\mathbf{k}^{2}\left( \boldsymbol{\theta }%
\cdot \boldsymbol{\mathcal{B}}\right) +\left( \mathbf{k}\cdot \boldsymbol{%
\mathcal{B}}\right) \left( \mathbf{k}\cdot \boldsymbol{\theta }\right) %
\right] .
\end{equation*}%
This means again that birefringence is absent up to the adopted accuracy of $%
o(\theta ^{2}).$ The positive branch of $k^{0}$ is 
\begin{equation}
k^{0}=-\left( \mathbf{k}\cdot \left( \boldsymbol{\theta }\times \boldsymbol{%
\mathcal{E}}\right) \right) +\left[ \left( 1-\boldsymbol{\theta }\cdot 
\boldsymbol{\mathcal{B}}\right) \left\vert \mathbf{k}\right\vert +\frac{1}{%
\left\vert \mathbf{k}\right\vert }\left( \mathbf{k}\cdot \boldsymbol{%
\mathcal{B}}\right) \left( \mathbf{k}\cdot \boldsymbol{\theta }\right) %
\right] .  \label{k0}
\end{equation}%
The group velocity is%
\begin{equation}
\mathbf{v}_{\mathrm{gr}}\equiv \frac{dk_{0}}{d\mathbf{k}}=\frac{\mathbf{k}}{%
\left\vert \mathbf{k}\right\vert }\left( 1-\left( \boldsymbol{\theta }\cdot 
\boldsymbol{\mathcal{B}}\right) \boldsymbol{\mathcal{-}}\frac{\left( \mathbf{%
k}\cdot \boldsymbol{\mathcal{B}}\right) \left( \mathbf{k}\cdot \boldsymbol{%
\theta }\right) }{\mathbf{k}^{2}}\right) +\boldsymbol{\theta }\frac{\left( 
\mathbf{k}\cdot \boldsymbol{\mathcal{B}}\right) }{\left\vert \mathbf{k}%
\right\vert }+\boldsymbol{\mathcal{B}}\frac{\left( \mathbf{k}\cdot 
\boldsymbol{\theta }\right) }{\left\vert \mathbf{k}\right\vert }-\left( 
\boldsymbol{\theta }\times \boldsymbol{\mathcal{E}}\right) ,  \label{vgr}
\end{equation}%
and its norm is 
\begin{eqnarray}
v_{\mathrm{gr}} &=&1-\left( \boldsymbol{\theta }\cdot \boldsymbol{\mathcal{B}%
}\right) \boldsymbol{\mathcal{+}}\text{ }\frac{\left( \mathbf{k}\cdot 
\boldsymbol{\mathcal{B}}\right) \left( \mathbf{k}\cdot \boldsymbol{\theta }%
\right) }{\mathbf{k}^{2}}-\frac{\left( \mathbf{k}\cdot \left( \boldsymbol{%
\theta }\times \boldsymbol{\mathcal{E}}\right) \right) }{\left\vert \mathbf{k%
}\right\vert }=  \notag \\
&=&1-\frac{\mathbf{k}_{\perp }^{2}}{\mathbf{k}^{2}}\theta _{3}\mathcal{B}%
_{3}+\frac{k_{2}k_{3}}{\mathbf{k}^{2}}\theta _{2}\mathcal{B}_{3}+\frac{1}{%
\left\vert \mathbf{k}\right\vert }\left( k_{3}\theta _{2}\mathcal{E}%
_{1}+k_{1}\theta _{3}\mathcal{E}_{2}-k_{2}\theta _{3}\mathcal{E}_{1}\right) .
\label{eq:group-velocity-2}
\end{eqnarray}%
Since $\boldsymbol{\theta }$ and $\boldsymbol{\mathcal{B}}$ are
pseudovectors, while $\boldsymbol{\mathcal{E}}$ is a vector, Eqs. (\ref{k0}%
), (\ref{vgr}) are vectors and (\ref{eq:group-velocity-2}) is a scalar.

Special cases can be obtained from the above by removing field components.
For instance, by setting $\boldsymbol{\mathcal{E}}=0$, one obtains the
following modification 
\begin{equation}
v_{\mathrm{gr}}=1-\frac{\mathbf{k}_{\perp }^{2}}{\mathbf{k}^{2}}\theta _{3}%
\mathcal{B}_{3}+\frac{k_{2}k_{3}}{\mathbf{k}^{2}}\theta _{2}\mathcal{B}_{3}
\label{eq:group-velocity-3}
\end{equation}%
of the group velocity (\ref{eq:group-velocity-1}), valid for the special case%
\footnote{%
The dispersion law in this special case was obtained in \cite{Guralnik2001}} 
$\theta _{2},\theta _{3},\mathcal{B}_{3}\neq 0,\,\boldsymbol{\mathcal{E}}=0.$
This case can be specified in an invariant manner by adding the invariant
conditions $\mathcal{F}\widetilde{\theta }=0$, $\left( F\theta \tilde{\theta}%
\right) =0$ and $\mathfrak{F}>0$, which have the effect of removing the
electric field from the special frame. Because of the factor of $k_{3}$ in
the $\theta _{2}$-term in (\ref{eq:group-velocity-3}), the propagation
transverse to the magnetic field still gives $v_{\mathrm{gr}}^{\perp
}=1-\theta _{3}\mathcal{B}_{3},$ which may be greater than unity, like in (%
\ref{eq:group-velocity-1}), but it also remains equal to unity for
propagation parallel to the magnetic field. This is no longer the case for
the more general Eq. (\ref{eq:group-velocity-2}): for the parallel
propagation, too, the group velocity (\ref{eq:group-velocity-2})

\begin{equation}
v_{\mathrm{gr}}=1+\theta _{2}\mathcal{E}_{1}  \notag
\end{equation}%
may exceed unity.

One can proceed in a similar fashion by specifying another particular case
of interest, by setting $\boldsymbol{\mathcal{B}}=0$. Then, by rotating the
coordinate system around the third axis we can annihilate the component $%
\mathcal{E}_{2}$ of the electric field, and annihilate the component $\theta
_{3}$ by the subsequent rotation around the first axis. Therefore, by
setting $B_{3}=$\ $\mathcal{E}_{2}=\theta _{3}=0$ (\ref{eq:group-velocity-2}%
) we get the following expression 
\begin{equation*}
v_{\mathrm{gr}}=1+\theta _{2}\mathcal{E}_{1}\frac{k_{3}}{\left\vert \mathbf{k%
}\right\vert }
\end{equation*}%
for the group velocity in an electric field directed along the first axis.
The photon propagating in the plane spanned by the mutually orthogonal
vectors of the electric field $\boldsymbol{\mathcal{E}}$ and the
noncommutativity vector $\boldsymbol{\theta }$ does so with unit speed (that
of light in the vacuum). For other directions it exceeds unity when the
three vectors $\boldsymbol{\mathcal{E}}$, $\boldsymbol{\theta }$ and $%
\mathbf{k}$ make a right triad, $\mathcal{E}_{i}=a\epsilon _{ijk}\theta
_{j}k_{k}$ with $a>0$.

\section{Concluding remarks}

We have considered, in the lowest order of the noncommutativity parameter,
photon propagation in an anisotropic medium, equivalent to the vacuum in the
space-space noncommutative electrodynamics with external constant electric
and magnetic fields. The most general case considered includes mutually
perpendicular electric and magnetic fields, arbitrarily oriented with
respect to the noncommutativity vector, taken together or -- as special
cases -- separately. Our consideration is based on the Lorentz-covariant
formalism of the Lorentz-non-invariant theory, that operates with the local
action written as a Lorentz-scalar including the noncommutativity tensor and
the background constant electromagnetic fields, violating Lorentz
invariance. We found the dispersion laws and derived expressions for the
group velocities. In neither case the birefringence occurs within the
accuracy adopted, i.e. the solutions of dispersion equations for two
propagating modes coincide. In neither case the direction of the
wave-vector, the photon 3-momentum, coincides with the direction of the
energy-momentum propagation given by the group velocity (see \cite{ChaSha}
for the Lorentz-non-invariant case and nonsymmetric energy-momentum tensor).
The direction of the wave-vector should not be referred to as the direction
of propagation. Also no conclusions on the speed of propagation can be
arrived at based on the phase velocity, contrary to what some authors are
inclined to do. However, the group velocity should be considered as carrying
information. Consequently it must not exceed the speed of light in vacuum, $%
c=1,$ without conflicting with the causality principle, even though we are
dealing with a Lorentz-non-invariant theory. (The known exception to this
rule made by the phenomenon of abnormal dispersion can be easily
circumvented by a redefinition of the group velocity for complex energy
instead of complex momentum \cite{ShabUs2010}). Contrary to the causality
requirement, we have found that the group velocity does exceed unity if
special relations between the background electromagnetic field and the
noncommutativity tensor are fulfilled. Moreover, in the general case
considered, there is no special direction of propagation relative to the
background fields that would exclude propagation with speed exceeding unity.
We consider this situation as a serious indication of inconsistency of the
theory.

What may the possible way out be?

In general physics courses, professors sometimes tell students that a
perfectly rigid body cannot be built, because the speed of sound in it would
be greater than the speed of light in vacuum. However, they do not explain
what mechanisms can prevent one from constructing such a body. A positive
example of such a mechanism may be found in quantum electrodynamics with
external fields, where a super-Plank background field leads to the
possibility of superluminal propagation. However, this field cannot be
achieved, because the instability destroying that field occurs earlier \cite%
{ShabUs2011}. Following that line we must admit that a certain mechanism
should exist that would exclude or forbid causality violating relations
between the background field and the noncommutativity tensor. Within the
simplest case of Section 4 this would mean a prohibition of the background
magnetic field being opposite to the noncommutativity vector. Such a
mechanism, however, is unknown; anyway it is not seen to be provided by the
field equations.

On the other hand, we may speculate that the presence of a superluminal
signal may not be considered a catastrophe, indeed, provided that the excess
over the speed of light is extremely small. We should take into account that
the realization of a time machine would require a Lorentz transformation
with speed $V<0$, enough to reverse the sign of the time coordinate. This
means that the inequality $|V|>1/v$ is at least necessary. (It is understood
here that the signal is superluminal, $v>1$, albeit the speed of the
reference frame does not exceed the speed of light, $|V|$ $<1$). Such must
be also the speed of the device that registers the arrival of the
superluminal signal and sends a superluminal signal back, as explained in 
\cite{dolgovnovikov}. Since the signal speed is expected to exceed unity
only just a little, the speed of that device should closely approach the
speed of light. Moreover, to get a sufficiently negative time interval, as
it is desirable for achieving a sufficiently remote past, it is required
that $|V|$ approach unity still closer. Since we never experimented with
such devices, we cannot state that the whole manifold of the established
physical facts contradicts the possibility of constructing a time machine
once we have at our disposal a superluminal signal, whereas the logical
paradox implied by this device is not alone sufficient for ruling it out.

Let us estimate the necessary speed of the detector/emitter in the case of
noncommutative electrodynamics with external fields considered in the
present paper. According to the present results, the excess of the group
velocity over unity $\Delta v=v^{\mathrm{gr}}-1$ is of the order of $\theta
B.$ It makes sense to take for the magnetic field its largest value known
from pulsars and magnetars, which is of the order of magnitude of
Schwinger's characteristic value $\frac{m^{2}}{e}$ or two orders higher.
With $m$ and $e$ being the electron mass and charge this makes approximately 
$4.4\times 10^{13}G$. According to the strongest estimate of the
noncommutativity parameter found in \cite{AdoGitShabVas}, $\theta <\left(
1000\mathrm{Tev}\right) ^{-2}$, then $\Delta v=\theta B<\sim 10^{-18}$.
Hence the speed of the emitter/detector in the pulsar magnetosphere must be
greater than $1-10^{-18}$ One cannot fathom this speed for a macroscopic
body. This corresponds to the speed of a cosmic proton with energy of $%
10^{18}$ev. The kinetic energy of the emitter/detector weighing one gram
would be $10^{41}ev$.

\section*{Acknowledgements}

Gitman thanks CNPq and FAPESP for permanent support. The work of A. Shabad
and D. Gitman is also partially supported by the Tomsk State University
Competitiveness Improvement Program, by a grant from \textquotedblleft The
Tomsk State University Academician D.I. Mendeleev Fund
Program\textquotedblright\ and also by RFBR under the Project 15-02-00293a.
A. Shabad also thanks USP for kind hospitality extended to him during his
stay in São Paulo, Brazil, and acknowledges the support of FAPESP under
Processo 2014/08970-1.

\section*{Appendix}

We follow the conventions adopted in \cite{shabtrudy}, in particular, the
metric is given by $\eta=\mathrm{diag}\left(1,1,1,-1\right)$, $%
\mu,\nu,\rho,...=1,2,3,0$ and $\varepsilon_{1230}=1$. We use the following
conventions with regard to contractions:

\begin{align*}
& F_{\alpha \beta }^{n}\equiv F_{\alpha }^{\,\,\alpha _{1}}F_{\alpha
_{1}}^{\,\,\alpha _{2}}...F_{\alpha _{n-1}\beta },\quad (\theta
F^{n})_{\alpha \beta }\equiv \theta _{\alpha }^{\,\,\alpha _{1}}F_{\alpha
_{1}\beta }^{n},\quad (\theta K)=(\theta K)_{\alpha }^{\,\,\alpha },\quad
n=0,1,2,3..., \\
& (Kk)_{\alpha }\equiv K_{\alpha \beta }k^{\beta }\,,\,\,(kK)_{\alpha
}\equiv k^{\beta }K_{\beta \alpha }\,\,,\,\,(Kk)_{\alpha }=-\left( kK\right)
_{\alpha }\,\,\mathrm{for}\,\,K^{T}=-K.
\end{align*}%
Thus, $\left( kF\theta k\right) \equiv k^{\mu }F_{\mu \nu }\theta ^{\nu
\sigma }k_{\sigma }$ and $\left( \theta F\right) =\left( F\theta \right)
\equiv \theta _{\alpha \beta }F^{\beta \alpha }$. We also have 
\begin{equation*}
\mathfrak{F}=-\frac{1}{4}F^{2}=\frac{1}{2}\left( \mathbf{B}^{2}-\mathbf{E}%
^{2}\right) \,,\,\,\mathfrak{G}=-\frac{1}{4}(F\tilde{F})=\mathbf{E}\cdot 
\mathbf{B}\,,
\end{equation*}%
where $\tilde{F}_{\mu \nu }=\frac{1}{2}\varepsilon _{\mu \nu \rho \sigma
}F^{\rho \sigma }$, $E^{i}=F^{i0}$ and $B^{i}=\frac{1}{2}\varepsilon
^{ijk}F^{jk}$. \ We also note the useful relations 
\begin{equation*}
\left( F\tilde{F}\right) _{\mu \nu }=-\eta _{\mu \nu }\mathfrak{G}%
\,,\,\,F_{\mu \nu }^{3}=-2\mathfrak{F}F_{\mu \nu }-\mathfrak{G}\tilde{F}%
_{\mu \nu },
\end{equation*}%
\begin{equation*}
\left( \tilde{F}^{2}\right) _{\mu \nu }=\frac{1}{2}\eta _{\mu \nu
}F^{2}-\left( F^{2}\right) _{\mu \nu }.
\end{equation*}%
We list some identities for the case where in the special frame one has $%
\theta _{0i}=0.$ The dual noncommutativity tensor is defined as $\tilde{%
\theta}_{\mu \nu }=\frac{1}{2}\varepsilon _{\mu \nu \rho \sigma }\theta
^{\rho \sigma }.$ The noncommutativity vector is $\theta _{i}=\epsilon
_{ijk}\theta _{jk},$ and its dual is $\tilde{\theta}_{i}=\epsilon _{ijk}%
\tilde{\theta}_{jk}.$ The following relations hold

\begin{align}
\left( \theta F\right) & =-2\boldsymbol{\theta }\cdot \mathbf{B,}  \notag \\
\left( kF^{2}k\right) & =-\left( k^{0}\right) ^{2}\mathbf{E}^{2}-2k^{0}%
\mathbf{k}\cdot \left( \mathbf{E}\times \mathbf{B}\right) +\left( \mathbf{k}%
\cdot \mathbf{E}\right) ^{2}-\mathbf{k}^{2}\mathbf{B}^{2}+\left( \mathbf{k}%
\cdot \mathbf{B}\right) ^{2},  \notag \\
\left( k\theta Fk\right) & =k^{0}\mathbf{k}\cdot \left( \boldsymbol{\theta }%
\times \mathbf{E}\right) -\mathbf{k}^{2}\left( \boldsymbol{\theta }\cdot 
\mathbf{B}\right) +\left( \mathbf{k}\cdot \mathbf{B}\right) \left( \mathbf{k}%
\cdot \boldsymbol{\theta }\right) ,  \notag \\
\left( kF^{2}\theta Fk\right) & =\left[ \left( k^{0}\right) ^{2}\mathbf{E}%
^{2}-\left( \mathbf{k}\cdot \mathbf{E}\right) ^{2}+2k^{0}\mathbf{k}\cdot
\left( \mathbf{E}\times \mathbf{B}\right) +\mathbf{k}^{2}\mathbf{B}%
^{2}-\left( \mathbf{k}\cdot \mathbf{B}\right) ^{2}\right] \left( \boldsymbol{%
\theta }\cdot \mathbf{B}\right) -\left( k^{0}\right) ^{2}\left( \mathbf{E}%
\cdot \mathbf{B}\right) \left( \boldsymbol{\theta }\cdot \mathbf{E}\right) 
\notag \\
+& \left( \mathbf{k}\cdot \mathbf{E}\right) \left( \mathbf{E}\cdot \mathbf{B}%
\right) \left( \mathbf{k}\cdot \boldsymbol{\theta }\right) -k^{0}\left( 
\mathbf{E}\cdot \mathbf{B}\right) \mathbf{k}\cdot \left( \boldsymbol{\theta }%
\times \mathbf{B}\right) .  \label{useful-ids}
\end{align}

\end{document}